\documentclass[aps,pre,twocolumn,bibnotes]{revtex4-1}
\usepackage{amsmath,amsthm,amsfonts,amssymb,bm}
\usepackage{graphicx,subfigure}
\usepackage{color}
\usepackage{natbib}
\usepackage{mciteplus}
\newcommand{\citeS}[1]{\cite{#1}}

\newcommand{\precis}{pr\'{e}cis }
\newcommand{\considere}{Consid\`{e}re }

\newcommand{\gdot}{\dot{\gamma}}

\newcommand{\be}{\begin{equation}}
\newcommand{\ee}{\end{equation}}
\newcommand{\beqna}{\begin{eqnarray}}
\newcommand{\eeqna}{\end{eqnarray}}

\newcommand{\sigmae}{\sigma_{\rm E}}
\newcommand{\edot}{\dot{\epsilon}}

\newcommand{\taur}{\tau_R}
\newcommand{\taud}{\tau_d}

\begin{document}

\title{Triggers and signatures of shear banding in steady and
  time-dependent flows}

\author{S. M. Fielding} 

\affiliation{Department of Physics, Durham University, Science
  Laboratories, South Road, Durham, DH1 3LE, UK}
  
\date{\today}
\begin{abstract} {This \precis is aimed as a practical field-guide to
    situations in which shear banding might be expected in complex
    fluids subject to an applied shear flow. Separately for several of
    the most common flow protocols, it summarises the characteristic
    signatures in the measured bulk rheological signals that suggest
    the presence of banding in the underlying flow field.  It does so
    both for a steady applied shear flow, and for the time-dependent
    protocols of shear startup, step stress, finite strain ramp, and
    large amplitude oscillatory shear.  An important message is that
    banding might arise rather widely in flows with a strong enough
    time-dependence, even in fluids that do not support banding in a
    steadily applied shear flow. This suggests caution in comparing
    experimental data with theoretical calculations that assume a
    homogeneous shear flow. In a brief postlude, we also summarise
    criteria in similar spirit for the onset of necking in extensional
    filament stretching.}

\end{abstract}

\maketitle

\section{Introduction}

\begin{figure*}[tbp]
\includegraphics[width=16.0cm]{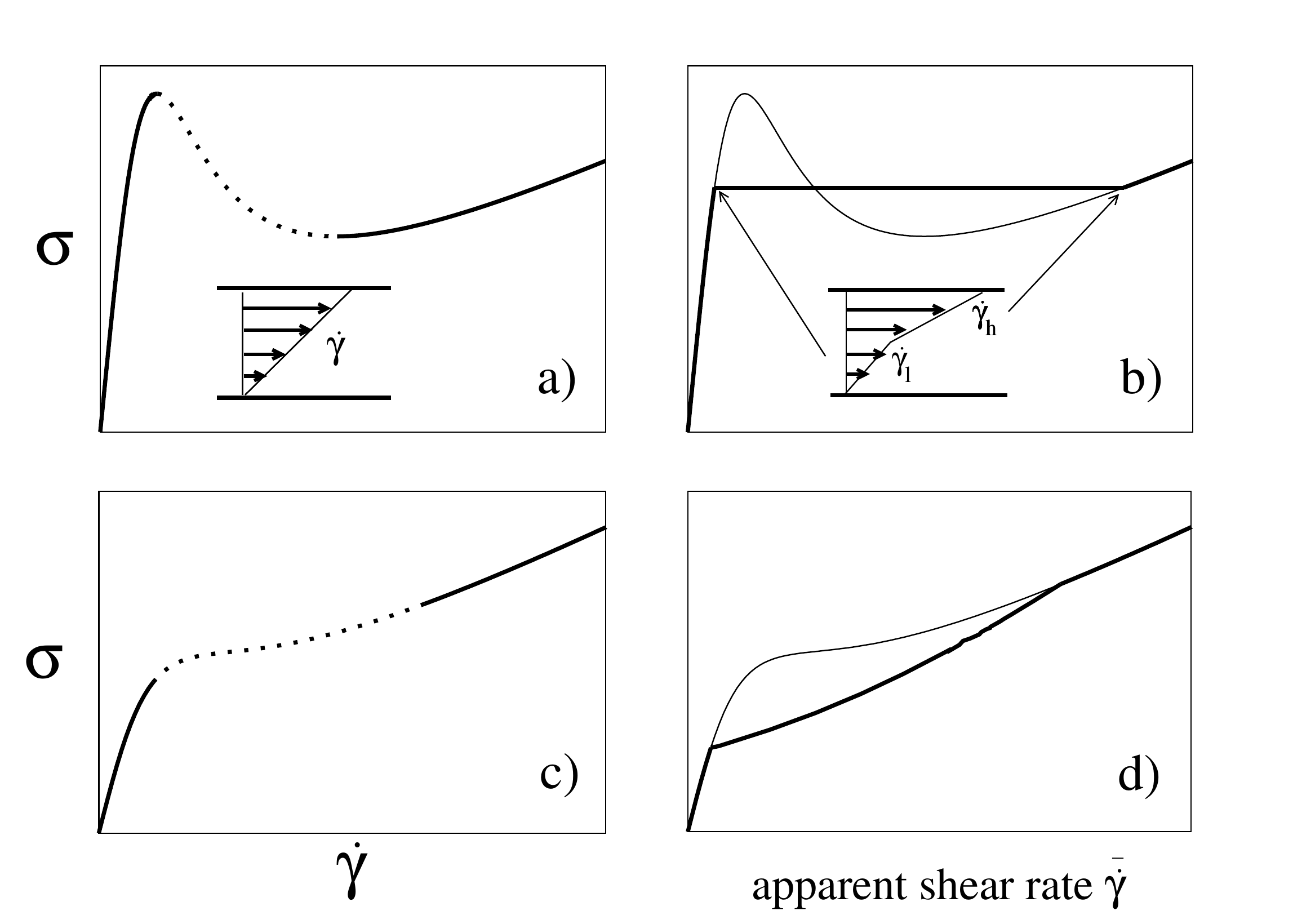}
\caption{Triggers and signatures of shear banding in a steadily
  applied shear flow. Left panels (a,c) show underlying constitutive
  relations between shear stress and shear rate, calculated within the
  assumption of a homogeneous shear flow.  In (a), a state of
  initially homogeneous shear flow is linearly unstable, in the regime
  of negative constitutive slope, to the formation of shear bands. The
  steady state flow curve (b) then has a characteristic stress plateau
  in the shear banding regime. The presence of flow-concentration
  coupling would extend the window of linear instability in (a). It
  can also render a purely monotonic constitutive curve linearly
  unstable to banding, as shown in panel (c). The signature of
  concentration coupling in the ultimate banded state is then an
  upward slope in the stress `plateau', as shown in panel (d). Line
  key: in the underlying constitutive curves (a) and (c) the thick
  solid lines denote homogeneous flow states that are linearly (though
  not necessarily absolutely) stable against shear banding, while the
  thick dotted lines denote homogeneous flow states that are linearly
  unstable to the formation of shear bands. In the flow curves (b) and
  (d) the thick solid lines represent steady flowing states (which are
  shear banded in some regimes as described above) while the thin
  solid lines represent the underlying constitutive curves, copied
  from (a) and (c).}
\label{fig:steady}
\end{figure*}

Many complex fluids show shear banding, in which a state of initially
homogeneous shear flow becomes unstable to the formation of coexisting
bands of differing shear rate, with layer normals in the flow-gradient
direction. See the sketches inset in Fig.~\ref{fig:steady}. (This
\precis concerns only this case of `gradient banding'; for a
discussion of `vorticity banding',
see~\citeS{ISI:000172419000010,ISI:000254405700003}.)  First observed
in wormlike micellar surfactants in the mid 1990s~\citeS{britton1997c},
it has since also been seen in lyotropic lamellar
phases~\citeS{Salmonetal2003a},
triblock copolymers~\citeS{Mannevilleetal2007a}, star
polymers~\citeS{rogers2008}, carbopol gel~\citeS{divoux2010},
clays~\citeS{martin-sm-8-6940-2012,Coussotetal2002c},
emulsions~\citeS{Coussotetal2002c} and (subject to ongoing
controversy~\citeS{ISI:000324679100008,ISI:000339141600010}) entangled
monodisperse linear
polymers~\citeS{ISI:000286154200002,ISI:000254645200053}.  For reviews,
see~\citeS{recentReview,ISI:000254405700003,ISI:000254405700004,ISI:000344142000001}.

To date, the majority of studies have focused on the case of steadily
applied shear flow, with banding as the ultimate steady state
response. However the last 5-10 years have seen a growing realisation
that banding might also arise rather widely in flow protocols with a
strong enough time-dependence, even in fluids that do not support
banding in steady
shear~\citeS{ISI:000315141600016,
ISI:000329357400005,
ISI:000263389500067,
ISI:000286879900011,
ISI:000344142000001,
PhysRevE.76.056106,
Manningetal2009a,
ISI:000285583500029,
ISI:000293534200007}.

In startup of steady simple shearing flow (``shear startup''), for
example, the (near ubiquitous) presence of an overshoot in the shear
stress startup signal has been identified as a possible trigger for
the formation of shear bands, at least transiently, en route to a
steady flowing
state~\citeS{ISI:000286879900011,ISI:000263389500067,ISI:000293534200007,ISI:000344142000001,ISI:000329357400005,ISI:000253331200055,Huetal2008a,ISI:000254645200053,Wangetal2009a,hu2007cre,Wangetal2009d,divoux2010,ISI:000295085700080,Martinetal2012a,Likhtmanetal2012,ISI:000246210200042,ISI:000315141600016,PhysRevE.76.056106,Manningetal2009a,ISI:000285583500029,ISI:000316969500006,Kurokawa}.
A declining time-dependent viscosity has been similarly identified as
a trigger for banding following the imposition of a step
stress~\citeS{ISI:000315141600016,Wangetal2009a, Huetal2008a,
  Wangetal2008c, Huetal2005a, Huetal2010a,
  Wangetal2009d,ISI:000294447600069,ISI:000280140800011,ISI:000329357400005,ISI:000344142000001}.
In these two protocols the time-dependence is transient, persisting
typically for a few strain units as a steady flow is established out
of an initial rest state. Accordingly, any bands must themselves be
transient and heal back to homogeneous flow in the final steady state
(unless the fluid also has banding as its ultimate steady state
response).  In soft `glassy' materials with sluggish relaxation
timescales, however, these startup bands might persist long enough to
be mistaken for the ultimate flow response of the material for any
practical purpose, despite being technically
transient~\citeS{ISI:000286879900011,ISI:000344142000001}.

Other flow protocols have sustained time-dependence: large amplitude
oscillatory shear (LAOS) is a notable
example~\citeS{ISI:000297489400003}.  In the mindset of the previous
paragraph, one might intuitively view a strain-imposed LAOS experiment
(hereafter abbreviated as LAOStrain), in some range of frequencies at
least, as a repeating process of forward then backward startup runs.
Any banding associated with the response of the same fluid to startup
of steady shear flow might then be anticipated to recur in each half
cycle of LAOS.  Banding would then be an integral, sustained feature
of LAOS, even if the fluid would not support banding in steadily
applied shear~\citeS{ISI:000263389500067,Carter}.

The aim of this \precis is to summarise criteria for, and signatures
of, the onset of banding, separately for each flow
protocol~\citeS{ISI:000315141600016}. It is offered as a field guide to
situations in which banding might be expected in complex fluids and
soft solids.  An important by-product is also to suggest that banding
might arise quite generically in flows with a strong enough time
dependence, even in materials that don't support banding in steady
state.

%
%

For each flow protocol in turn, we give a
criterion~\citeS{ISI:000315141600016} for the onset of banding in terms
of a characteristic signature in the shape of the relevant bulk
rheological response function for that protocol, {\it e.g.}, stress
versus strain in shear startup. As a starting point for a hydrodynamic
stability calculation, this response function is first calculated for
a `base state' in which the flow is assumed to stay homogeneous. A
linear stability analysis then reveals the point at which this base
state first becomes unstable to banding, and gives an onset criterion
in terms of the functional shape of the response function associated
with that base state.  However because the flow is by definition
homogeneous before it bands, these onset criteria can also be applied
directly to the functional shape of the experimentally measured
rheological response function. (This concept is explained more fully
after (\ref{eqn:startup}) below.)

\begin{figure*}[tbp]
\includegraphics[width=16.0cm]{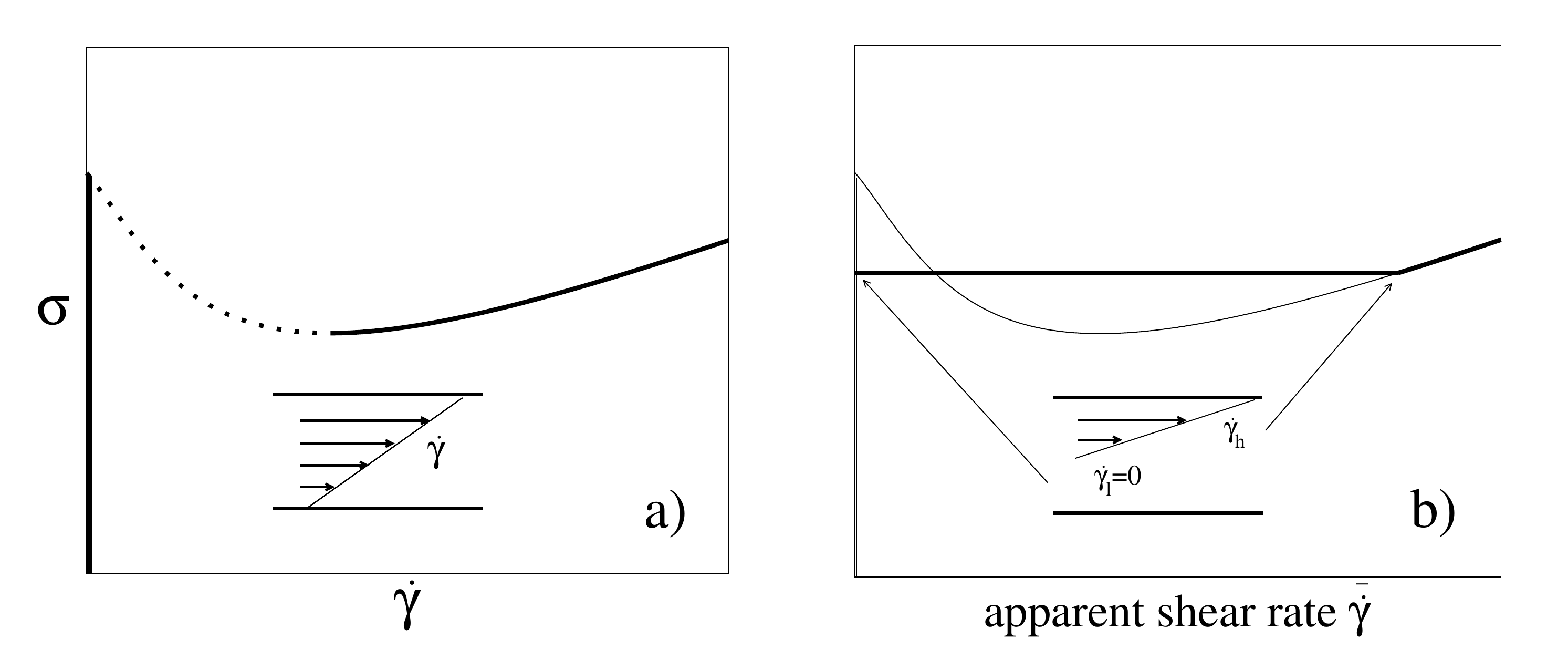}
\caption{Triggers and signatures of shear banding in a steadily
  applied shear flow in a yield stress fluid. Left panel (a) shows an
  underlying constitutive relation between shear stress and shear
  rate, calculated within the assumption of a homogeneous shear flow.
  A state of initially homogeneous shear flow is linearly unstable in
  the regime of negative constitutive slope to the formation of shear
  bands. The steady state flow curve (b) then has a characteristic
  stress plateau in the shear banding regime. Compared with the
  corresponding sketch for ergodic fluids in Fig.~\ref{fig:steady},
  the low-shear branch of the constitutive curve lies vertically up
  the stress axis in (a) and the corresponding band in (b) is
  unsheared. (As discussed in the text, concentration coupling is also
  possible in these yield stress materials, but we have not sketched
  it separately.) Line key: as in Fig.~\ref{fig:steady}.}
\label{fig:yield}
\end{figure*}

For anyone not wishing to read the rest of the paper, the signatures
are summarised at a glance for steady applied shear flow in
Figs.~\ref{fig:steady} and~\ref{fig:yield}. The signatures for the
transiently time-dependent flows of shear startup, step stress, and
finite strain ramp are likewise summarised in
Figs.~\ref{fig:startup},~\ref{fig:creep} and~\ref{fig:ramp}
respectively. For LAOS, with its sustained time-dependence, we sketch
in Fig.~\ref{fig:laos} the regions of the plane of applied strain rate
amplitude and cycle frequency in which banding is anticipated.

Once significant banding develops, it in general changes the shape of
the response function compared to that calculated within the
assumption of homogeneous flow. This provides a note of caution to the
endeavour of benchmarking new constitutive models by comparing
homogeneous calculations with experiment in any of the widespread
situations where banding might arise.

Most theoretical work to date in this area has been on models of
linear entangled polymeric fluids (polymer solutions and melts, and
wormlike micelles)~\citeS{ISI:000315141600016,ISI:000329357400005};
and of soft `glassy' materials (foams, dense emulsions, dense
colloids, microgels, {\it
  etc.})~\citeS{ISI:000286879900011,ISI:000344142000001,PhysRevE.76.056106,Manningetal2009a,ISI:000285583500029},
which typically show a yield stress and rheological aging.  However it
is hoped that the criteria might apply universally. These two classes
are exemplary only, and were selected for study because they are the
most familiar to this author. Indeed this \precis is highly selective
and focused mainly on the author's own work. Work by others to further
generalise, or delineate the regimes of applicability of these criteria
would be very welcome.


%
%

\section{Shear banding in steady imposed flows}
\label{sec:steadily}

We consider first the long-time response of a fluid to a steadily
applied shear flow. In the interests of definite vocabulary, we shall
use the term ``constitutive curve'' to denote a material's underlying
stationary relation $\sigma(\gdot)$ between shear stress and shear
rate, calculated within the assumption that the flow remains spatially
uniform with the shear rate everywhere equal to $\gdot$.  Although
stationary, however, states on this curve may not be stable against
banding. Where shear bands form, we term the steady state relation
between shear stress and shear rate the composite ``flow curve'', with
the relevant shear rate now being the spatially averaged value across
the cell, {\it i.e.}, the relative wall velocity normalised by the gap
size, often termed the ``apparent shear rate''. In the absence of
banding, these two curves coincide.

\subsection{Steady state bands}
\label{sec:steady}

A state of initially homogeneous shear flow is known to be linearly
unstable in any regime where the fluid's underlying constitutive curve
has negative slope~\citeS{Yerushalmi70}:
\be
\label{eqn:steady}
\frac{d\sigma}{d\gdot} <0.
\ee
See Fig.~\ref{fig:steady}a. Shear bands then form, and the steady
state composite flow curve displays a characteristically flat
plateau~\citeS{PhysRevLett.71.939}, Fig.~\ref{fig:steady}b. In a
curved flow cell, this plateau will in fact have a slight positive
slope~\citeS{ISI:000085655200005}. (Indeed, taking into account
intrinsic heterogeneity in the flow field due to device curvature is
an important step in benchmarking constitutive models, even in the
absence of true banding.) In the windows of shear rate within the
steady state banding regime, but either side of the linearly unstable
regime, an initially homogeneous flow is metastable to
banding~\citeS{ISI:A1997XQ47400002}. In this regime, in a slow strain
rate sweep at least, a finite amplitude perturbation to an initially
homogeneous flow is required to initiate banding. Possible sources
include initial heterogeneities following sample preparation,
mechanical noise in the rheometer, or true thermal noise. (In a shear
startup at such shear rates, a linear instability can arise during
that time-dependent startup process. However we defer further
discussion of these time-dependent phenomena to Sec.~\ref{sec:startup}
below.)

In two-component viscoelastic fluids (solutions), spatial variations
in the flow field are in general dynamically coupled to variations in
the concentration field
$\phi$~\citeS{schmitt95,brochard-m-10-1157-1977,HelfFred89,DoiOnuk92,Miln93,wu-prl-66-2408-1991,beris94,SBJ97}.
This provides a positive feedback mechanism that enhances a fluid's
tendency to form shear
bands~\citeS{ISI:000185756400080,ISI:000183381200013,ISI:000184319400008},
giving a modified onset criterion of the general form
\be 
\label{eqn:phi}
\frac{d\sigma}{d\gdot} + C_{\gdot\phi} < 0.  
\ee
In this inequality, $C_{\gdot\phi}$ is a flow-concentration coupling
term. Its full form is given in Eqn. (4.20) of
Ref.~\citeS{ISI:000185756400080} and is rather complicated. However in
essence its numerator comprises the derivative of shear stress with
respect to concentration, multiplied by the derivative of a normal
stress-like variable with respect to shear rate. Its denominator
comprises (in essence) the (bare) osmotic modulus minus the derivative
of a normal stress-like variable with respect to concentration.  The
overall effect of this coupling is such that the regime in which an
initially homogeneous flow is predicted to be linearly unstable to the
formation of shear bands extends slightly beyond the dotted region in
Fig.~\ref{fig:steady}a).

More importantly, flow-concentration coupling can also render a weakly
sloping but monotonic constitutive curve unstable to banding.  See
Fig.~\ref{fig:steady}c).  This was first explored in the context of
polymeric
fluids~\citeS{ISI:000185756400080,ISI:000183381200013,ISI:000184319400008}.
It has recently been studied again in
polymers~\citeS{ISI:000320001200003,ISI:000341175200016}, and also
applied to yield stress colloidal
fluids~\citeS{ISI:000286754700021,ISI:000345090400010}.  The signature
of concentration coupling in the steady state composite flow curve is
an upward slope in the `plateau' of the banding regime.  For strong
coupling this can be quite pronounced, as sketched in
Fig.~\ref{fig:steady}d). For weak coupling it may go unnoticed.

The sketches in Figs.~\ref{fig:steady} pertain to ergodic fluids with
fixed, finite stress relaxation timescales.  However shear banding
also arises widely in non-ergodic soft glassy
materials~\citeS{ISI:000344142000001}, which have a yield stress
$\sigma_{\rm Y}$ associated with sluggish and often aging stress
relaxation timescales.  (Throughout we use the term yield stress to
denote the limiting shear stress obtained as $\gdot\to 0$ in a slow
strain rate sweep down the flow curve.)  In this case the constitutive
curve's high viscosity branch lies vertically up the $\sigma$ axis, as
sketched in Fig.~\ref{fig:yield}.  The band associated with this
branch is then unsheared, and coexists with a flowing band of non-zero
shear rate on the other flow
branch~\citeS{Coussotetal2002c,ISI:000242151800010}, as sketched in
the inset to Fig.~\ref{fig:yield}b). Otherwise, the comments of this
section generally apply.

\subsection{Oscillatory and chaotic shear bands}
\label{sec:rheochaos}

The discussion so far has assumed that a fluid's ultimate response to
a steadily applied shear will be a state of steady flow.  In some
cases, however, oscillations or even chaotic fluctuations can
arise~\citeS{bandyopadhyay-prl-84-2022-2000,ganapathy-prl-96--2006}: not
transiently, but as the ultimate response of the material, sustained
as long as the flow remains applied. Spatio-temporally oscillating and
chaotic shear bands were explored in
Refs.~\citeS{ISI:000189266100020,ISI:000229044600017,aradian-pre-73--2006}.
We do not discuss them further here.
 
\section{Shear banding in transiently time-dependent flows}
\label{sec:transient}

We now turn to protocols in which the applied flow is itself
inherently time-dependent. In this section we consider situations in
which that time-dependence is transient in nature: arising either
during a process whereby a steady flow is established out of an
initial rest state (in shear startup or following the imposition of a
step stress), or after a finite strain ramp as the system relaxes back
to equilibrium. For the remainder of this \precis we ignore
concentration coupling, deferring to future work a study of its effects
in time-dependent flows.

\subsection{Shear startup}
\label{sec:startup}

%
\begin{figure}[tbp]
\includegraphics[width=8.0cm]{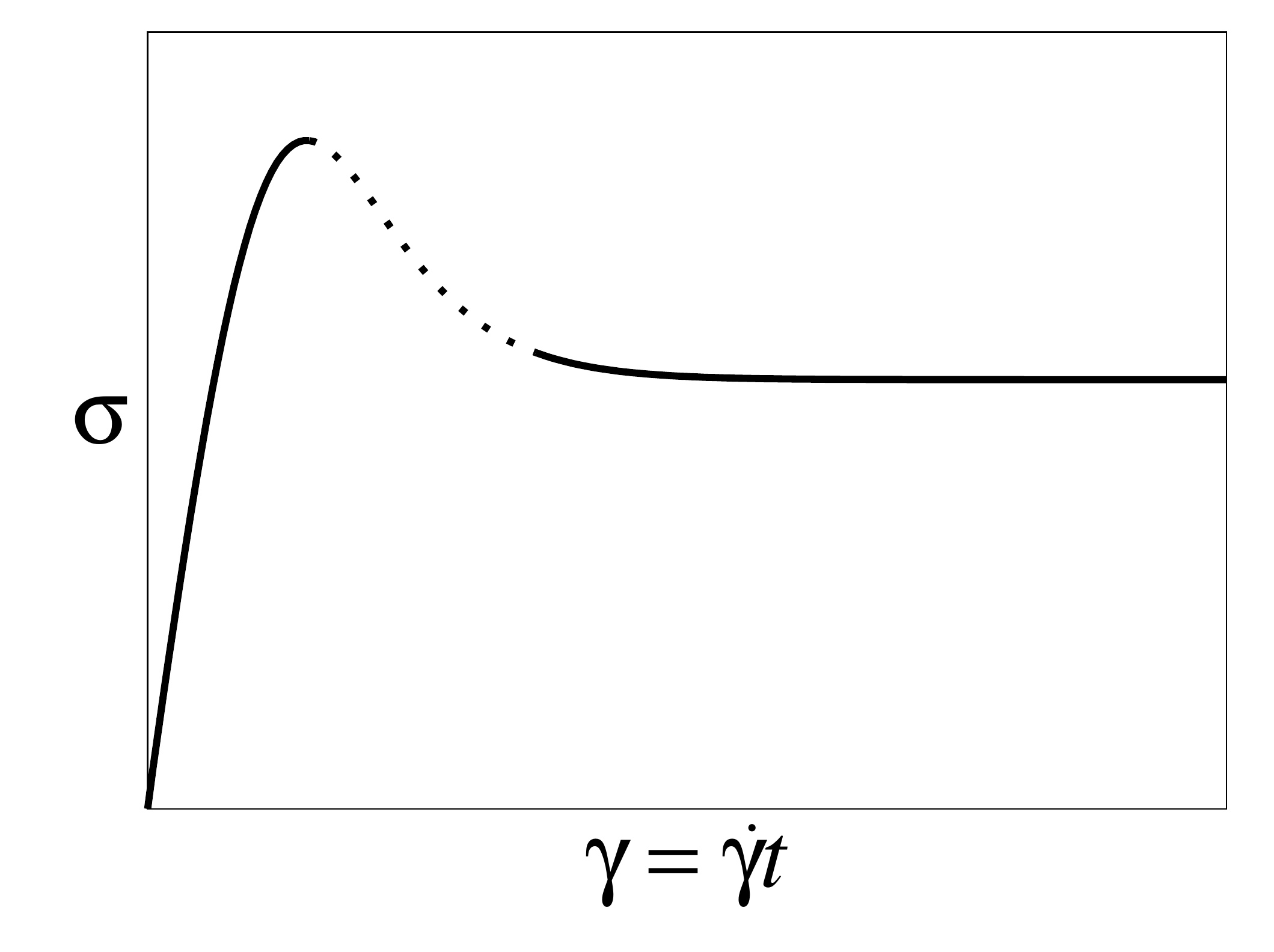}
\caption{Typical shear stress response in shear startup. The region of
  linear instability to the formation of shear bands is sketched as
  dotted. Depending whether the same fluid also supports steady state
  bands at the flow rate in question, according to the sketches in
  Fig.~\ref{fig:steady}, these startup bands either persist to steady
  state or heal back to homogeneous flow. (In the former case, the
  line should be dotted even at long times.)}
\label{fig:startup}
\end{figure}
%

A common flow protocol consists of taking a sample that is initially
at rest and with any residual stresses well relaxed, then at some time
$t=0$ suddenly jumping the strain rate to some value $\gdot$ that is
held constant thereafter.  Measured in response to this is the shear
stress startup signal $\sigma(t)$ as a function of the time $t$ (or
accumulated strain $\gamma=\gdot t$) since the inception of the flow.
This typically evolves as sketched in Fig.~\ref{fig:startup}, with an
initial regime of linear elastic response in which the stress rises
proportionally with the strain, followed by an overshoot at a strain
$\gamma = O(1)$, then a final decline to a steady state value on the
flow curve at the given strain rate.

In Ref.~\citeS{ISI:000315141600016}, it was argued that the presence of
an overshoot in this startup signal is generically indicative of a
strong tendency to form shear bands, at least transiently during the
startup process.  Accordingly -- though in sketch form only (we
discuss corrections and caveats below) -- the criterion for the onset
of banding in startup is
\be
\label{eqn:startup}
\frac{d\sigma}{d\gamma}<0,
\ee
as indicated by the dotted line in Fig.~\ref{fig:startup}. 

This criterion (\ref{eqn:startup}) was derived by first calculating
the stress signal associated with an underlying time-evolving
homogeneous base state startup flow, artificially imposing (for the
purposes of that preliminary calculation) the constraint that the flow
must remain homogeneous.  Performing a linear stability analysis for
the dynamics of heterogeneous fluctuations about this time-evolving
base state then shows that it first becomes unstable to the formation
of shear bands just after the stress overshoot. In this way, an
overshoot in the startup signal associated with that underlying
time-evolving base state is predicted to act as a trigger to banding.
These considerations can then be applied to real data by recognising
that before any banding arises in any given experiment, the flow is
(by definition) homogeneous and so accords with the base state of the
homogeneous calculation. We thus recognise that the criterion
(\ref{eqn:startup}) can also be applied directly to the experimentally
measured stress startup signal.

A common misconception is that it is instead the onset of banding that
causes the stress drop.  While it is true that once significant
banding develops it in general reduces the stress compared to that
calculated assuming homogeneity, thereby accentuating the drop, the
primary direction of causality (at least in all the models this author
has studied to date) is the opposite: the onset of banding is
triggered by the stress drop, not vice versa.

As noted above, this discrepancy between the stress signal of the
homogeneous base state and that of the full shear banded flow should
provide a note of caution to the common practice of benchmarking new
constitutive models by comparing experimental startup data with
calculations that assume the flow to remain homogeneous.

In any fluid for which the ultimate constitutive response also admits
steady state banding as in Sec.~\ref{sec:steady} above, these bands
that form during startup will persist to steady state. In fluids that
don't support steady state banding, the startup bands instead heal
back to homogeneous shear. Indeed, in this case the tendency to form
bands persists only transiently. With this in mind, it is important to
note that not only must condition (\ref{eqn:startup}) be satisfied,
but the banding instability must also be strong enough for long enough to
ensure that observable banding can develop before homogeneous flow is
recovered. Clearly, a more pronounced stress overshoot is more likely
to give rise to more strongly observable banding.

Despite technically being transient, however, in soft glasses with
sluggish relaxation timescales, the bands may persist long enough to
be mistaken for the ultimate flow response of the
material~\citeS{ISI:000286879900011,ISI:000344142000001}.  Soft
glasses are also predicted to exhibit a strong age dependence: a
sample that is older and more solid-like before the flow commences
shows a stronger
overshoot~\citeS{derec-pre-67--2003,rogers-jr-54-133-2010,ISI:000288162500031,ISI:000295085700080,ISI:000341025500004},
and is predicted to show more pronounced startup
bands~\citeS{ISI:000286879900011,ISI:000344142000001}.

Intuitively, then, the startup banding just described is triggered as
the material `yields' and starts to flow, post-overshoot.  However it
is important to note that while the stress drop and associated banding
may indeed arise from actual yielding, {\it i.e.}, increasing
plasticity, as in a soft
glass~\citeS{ISI:000286879900011,ISI:000344142000001}, it can equally
stem from falling elasticity. The latter scenario is
predicted~\citeS{ISI:000293534200007,ISI:000329357400005,ISI:000263389500067,ISI:000315141600016}
by the Rolie-poly model~\citeS{likhtmangraham03} of linear polymers at
shear rates exceeding the inverse reptation time $\taud$ but less than
the inverse chain stretch relaxation time $\taur$. (A small correction
to (\ref{eqn:startup}) in this context is however discussed below.) In
this regime the stress startup curve is a unique function of strain,
independent of strain rate, but nonetheless sufficiently nonlinear to
show an overshoot. Although the coincidence of the criterion
(declining stress as a function of accumulated strain) for banding
instability in both these scenarios of plastic yielding and
falling-elasticity is highly suggestive of common physics, further
work is needed to elucidate this fully.

Evidence to date for banding associated with stress overshoot in
startup can be summarised as follows.  It has been seen experimentally
in polymeric fluids including wormlike
micelles~\citeS{ISI:000253331200055,Huetal2008a} and linear
polymers~\citeS{ISI:000254645200053,Wangetal2009a,hu2007cre,Wangetal2009d};
and in soft glassy materials including carbopol
gel~\citeS{divoux2010,ISI:000295085700080}, Laponite
clay~\citeS{Martinetal2012a,gibaud2008}, a non-Brownian fused silica
suspension~\citeS{Kurokawa} and waxy crude
oil~\citeS{ISI:000341025500004}.  Molecular simulations have captured
it in polymers~\citeS{Likhtmanetal2012,Khomami}, a model colloidal
gel~\citeS{ISI:000342206200002} and molecular
glasses~\citeS{ISI:000246210200042,varnik-jcp-120-2788-2004,Shrivastav}.
A model foam displayed it
in~\citeS{ISI:000250675300003,ISI:000278158800014}. Linear stability
analysis and nonlinear simulations predict it in the Rolie-poly model
of
polymers~\citeS{ISI:000315141600016,ISI:000329357400005,ISI:000263389500067,ISI:000293534200007},
the soft glassy rheology (SGR) and fluidity models of soft
glasses~\citeS{ISI:000286879900011,ISI:000344142000001,ISI:000322544200022},
the STZ model of amorphous elastoplastic
solids~\citeS{PhysRevE.76.056106,Manningetal2009a,Falk}, a mesoscopic model
of plasticity~\citeS{ISI:000285583500029}, and a model of polymer
glasses~\citeS{ISI:000316969500006}

With the aim of providing a unified understanding of all these
observations, a theoretical criterion for the onset of banding in
startup was derived analytically in Ref.~\citeS{ISI:000315141600016}
on the basis of a constitutive model written in a highly generalised,
though still differential, form. It was shown to indeed be closely
associated with stress overshoot, consistent with the evidence
summarised in the previous paragraph. It also showed full quantitative
agreement with numerical calculations in the Rolie-poly
model~\citeS{ISI:000329357400005}.

However to make progress analytically the calculation allowed for only
two viscoelastic variables: the viscoelastic shear stress $\sigma_v$
and one component of normal stress $n$.  Specifically, for this case
of simple shear flow it considers a force balance condition for a
total shear stress $\sigma=\sigma_v+\eta\gdot$ comprising a
viscoelastic contribution $\sigma_v$ and a Newtonian solvent stress
$\eta\gdot$.  Generalised constitutive dynamics for the viscoelastic
stresses are then prescribed as $\dot{\sigma_v}=f(\gdot,\sigma_v,n)$
with $\dot{n}=g(\gdot,\sigma_v,n)$, with $n$ a normal stress variable.
The functions $f$ and $g$ are left unspecified in the interests of
generality, but include stress relaxation on a timescale $\tau$. More
generally still, however, more viscoelastic variables besides
$\sigma_v$ and $n$ should be included (as will usually arise after
extracting componentwise equations for a fully tensorial constitutive
model in shear). Examples include the second normal stress (even in a
single mode description); or contributions to the stress from
additional modes with faster relaxation times.

Accordingly, the status of (\ref{eqn:startup}) more generally remains
unclear: it does not appear to apply in a straightforward way in the
Giesekus model, for example~\citeS{ISI:000329357400005}.  However
(\ref{eqn:startup}) does correctly predict the onset of banding
instability during startup in the Rolie-poly
model~\citeS{ISI:000329357400005} (with a small correction discussed
below) and in models of soft
glasses~\citeS{ISI:000286879900011,ISI:000344142000001}, including the
SGR model (which, being of integral form, effectively has infinitely
many viscoelastic modes); and accords well with the evidence from
experiment and molecular simulation described above.

Taken together, then, the evidence to date for banding triggered by
overshoot appears widespread and quite convincing. It suggests that
experimentalists should be alert to possible banding in any startup
experiment where the stress signal shows a strong overshoot; and that
theorists should exercise caution in benchmarking homogeneous
calculations against experiment.

As just described, the criterion derived in
Ref.~\citeS{ISI:000315141600016} is closely associated with the
overshoot in the stress startup curve, as written in
(\ref{eqn:startup}).  In fact, the full formula (Eqn. 20 in the
Supplementary Material of Ref.~\citeS{ISI:000315141600016}) contains not
only the slope of the stress with respect to strain, but also a
smaller correction term involving the curvature: the instability
technically first sets in just before overshoot, as the stress signal
curves down after the initial regime of linear elasticity.  This
agrees fully with numerics in the Rolie-poly
model~\citeS{ISI:000329357400005}, though in practical terms only
modest bands with weakly differentiated shear rates arise before the
overshoot.

The discussion in this section has focused on a single startup
experiment in which the shear rate is discontinuously jumped from zero
to some constant value. Similar effects have also been explored
experimentally~\citeS{ISI:000313006100057}
in fast upward shear rate sweeps in soft
glasses.

\subsection{Step stress}
\label{sec:creep}

\begin{figure}[tbp]
\includegraphics[width=8.0cm]{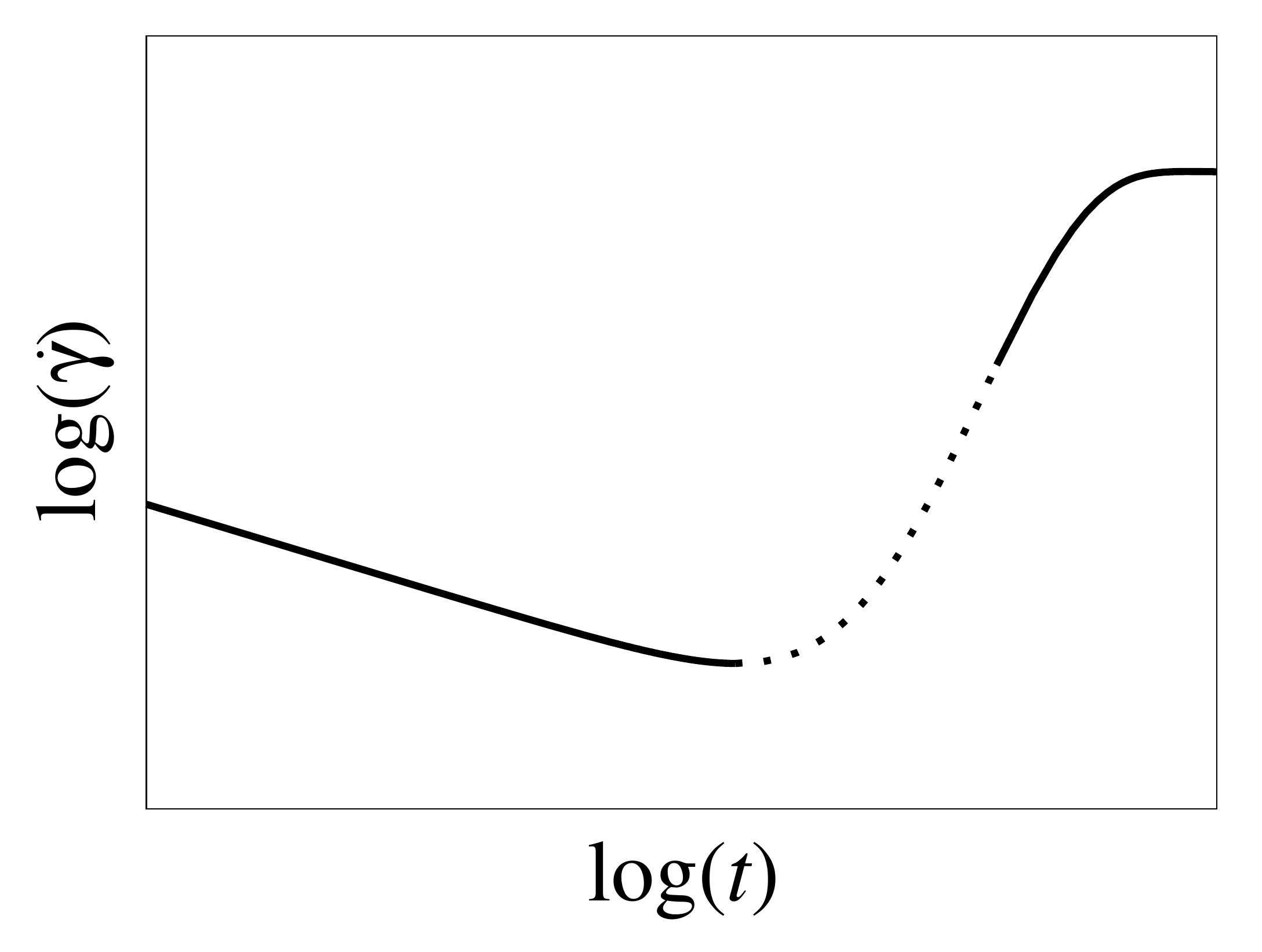}
\caption{Typical evolution of the time-differentiated creep response
  curve following the imposition of a step stress.  The regime of
  linear instability to banding is shown as dotted.}
\label{fig:creep}
\end{figure}

We consider now a previously undisturbed sample subject at some time
$t=0$ to the imposition of a shear stress that is held constant
thereafter.  Typically measured in response to this is the creep curve
$\gamma(t)$, often reported as its time-differential $\gdot(t)$. In
many cases this shows an initial regime of slow creep in which the
strain rate progressively declines, followed by a more rapid yielding
process in which the strain rate increases to attain its final steady
state on the flow curve. See Fig.~\ref{fig:creep}.

In~\citeS{ISI:000315141600016}, it was shown that the criterion for
instability to banding is that this differentiated creep curve obeys:
\be
\label{eqn:creep}
\frac{d^2\gdot}{d t^2}/\frac{d\gdot}{dt}>0.
\ee
A material is therefore predicted to be unstable to forming shear bands,
at least transiently, if its differentiated creep curve simultaneously
slopes upward and curves upward as a function of time. See the dotted
regime in Fig.~\ref{fig:creep}. (Simultaneous downward slope and curvature
are also predicted to initiate banding, but this author does not know
of any instances of such response.)

As in shear startup discussed above, this criterion is derived by
first calculating the creep response of an underlying base state in
which the sample is assumed to remain homogeneous, then performing a
linear stability analysis to determine the condition under which that
base state first becomes unstable to banding. And, by arguments
analogous to those just after (\ref{eqn:startup}), because the flow is
by definition homogeneous before it bands, (\ref{eqn:creep}) can also
be applied to the experimentally measured creep curve.

As in shear startup, then, banding is predicted to arise as the
material starts to `yield' towards a flowing state after a regime of
initially more solid-like response.  In the models of polymeric fluids
that this author has studied to date, such a scenario arises to most
pronounced effect at imposed stresses just above the local maximum in
a nonmonotonic constitutive curve of the form in
Fig.~\ref{fig:steady}a), or in the region of weak positive slope in a
monotonic curve as in
Fig.~\ref{fig:steady}c)~\citeS{ISI:000329357400005}.  In the soft
glassy rheology model, which has a monotonic constitutive curve, it
arises most strongly for imposed stresses just above the yield
stress~\citeS{ISI:000315141600016,ISI:000344142000001}, and is more
pronounced in a sample aged into a more solid-like state before the
stress is applied.

In all cases studied to date these bands heal back to homogeneous flow
in the ultimate steady state, consistent with the fact that steady
state banding can only be accessed under conditions of imposed strain
rate. (Recall that the flow curve is a flat function of strain rate in
the banding regime, at least in the absence of concentration
coupling.) In soft glasses with sluggish relaxation times, however,
they can persist a very long time, particularly for initially well
aged samples subject to imposed stresses only just exceeding the yield
stress~\citeS{ISI:000315141600016,ISI:000344142000001}.

%
%

Evidence to date for banding after a step stress can be summarised as
follows.  It has been seen experimentally in
polymers~\citeS{Wangetal2009a, Huetal2010a, Wangetal2009d}, wormlike
micelles~\citeS{Huetal2008a,Wangetal2008c, Huetal2005a}, carbopol
gel~\citeS{ISI:000294447600069,ISI:A1990DZ50400005}, carbon
black~\citeS{ISI:000280140800011,ISI:000332461800012} and a colloidal
glass~\citeS{ISI:000357577400001}. Particle based simulations of
molecular glasses have captured it~\citeS{ISI:000325376600001}. Linear
stability calculations and direct numerical simulations have
demonstrated it in the Rolie-poly and Giesekus models of
polymers~\citeS{ISI:000329357400005}, though as a weaker effect in the
latter model.  Stochastic simulations have confirmed it as a strong,
age-dependent phenomenon for imposed stresses just above the yield
stress in the soft glassy rheology (SGR)
model~\citeS{ISI:000315141600016,ISI:000344142000001}.

More universally, the criterion (\ref{eqn:creep}) was derived within a
constitutive model written in highly general, though still
differential form~\citeS{ISI:000315141600016}.  In contrast to its
counterpart (\ref{eqn:startup}) for startup, which is subject to the
caveat discussed in the previous section, the derivation of
(\ref{eqn:creep}) placed no limitations on the number of dynamic
viscoelastic variables present in the constitutive description.
Accordingly it should even apply to constitutive models of integral
form (which can be cast in differential form with infinitely many
dynamical variables). This is consistent with the observation of
banding following a step stress in the SGR model, which indeed has a
constitutive equation of integral
form~\citeS{ISI:000315141600016,ISI:000344142000001}.

On the basis of the evidence just summarised, we suggest that the
criterion (\ref{eqn:creep}) for instability to shear banding following
the imposition of a step stress might apply universally to all
materials. 

\subsection{Rapid finite strain ramp}
\label{sec:ramp}

\begin{figure}[tbp]
\includegraphics[width=8.0cm]{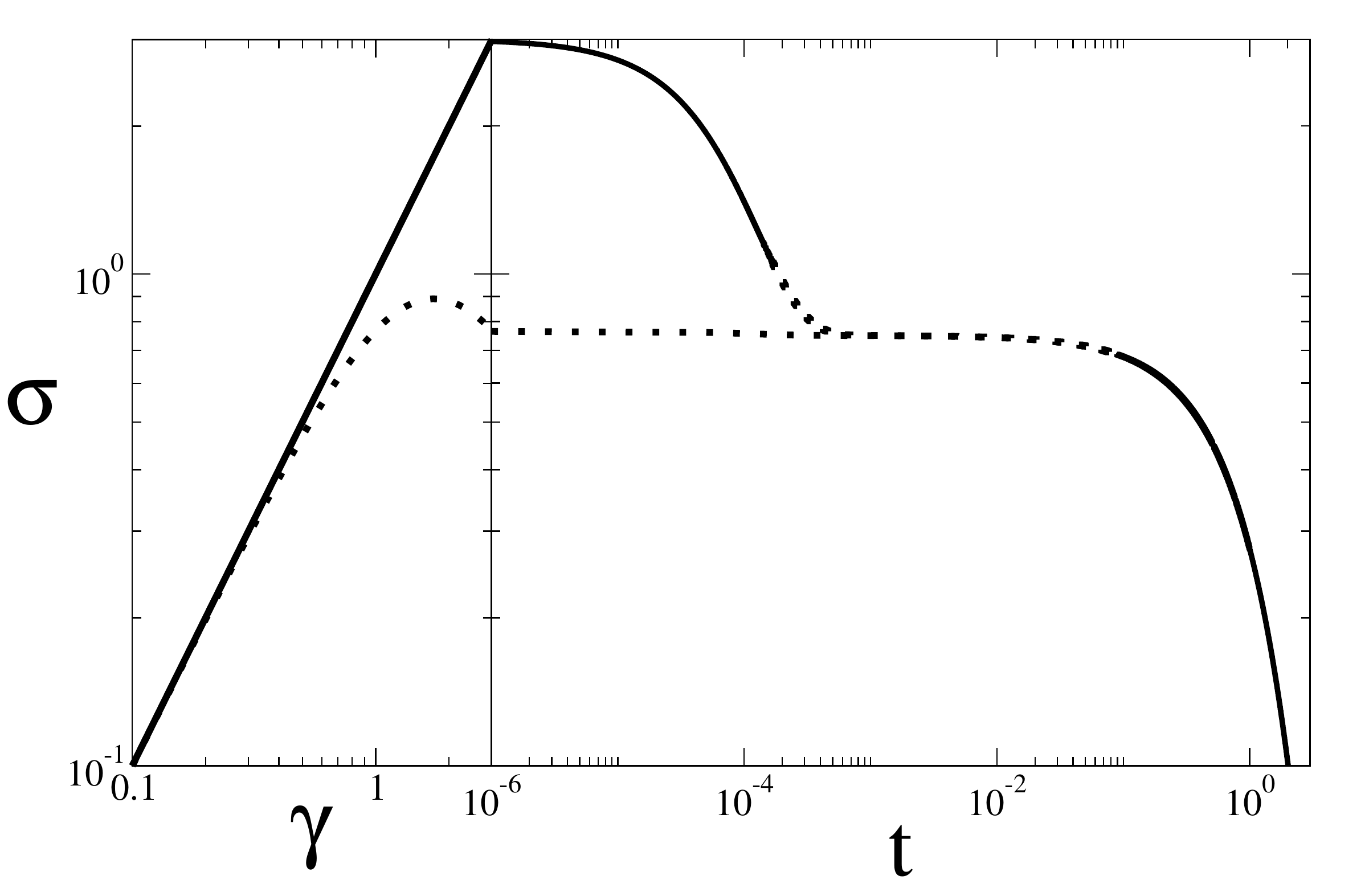}
\caption{Typical evolution of the shear stress with strain during a
  rapid strain ramp, then decay of the stress as a function of the
  time post-ramp. Regimes of linear instability to banding are
  sketched as dotted.  Data taken from calculations performed with the
  Rolie-poly model~\citeS{ISI:000329357400005}.}
\label{fig:ramp}
\end{figure}

We now turn to the protocol that is sometimes called `step strain',
but is in practice a fast finite strain ramp: a previously undisturbed
material is subject after some time $t=0$ to a linearly increasing
strain $\gamma=\gdot t$. Once some accumulated strain $\gamma_0$ is
reached the strain rate is set to zero, and the strain remains
constant at $\gamma=\gamma_0$. Measured in response is the stress as a
function of the time (or accumulated strain) during the ramp itself,
then the stress decay as a function of time as the system relaxes back
to equilibrium post-ramp. See Fig.~\ref{fig:ramp}.

%
%

During the ramp itself, the stability properties of an initially
homogeneous base state to the onset of banding are the same as in
shear startup, because the two protocols are the same in this regime.
However we focus here on ramps that most closely approximate the
notion of a `step strain', and are therefore sufficiently fast that no
meaningful banding has time to develop during the ramp (even if a
homogeneous shear is technically unstable to banding during it).  Our
interest is instead in whether any appreciable banding arises during
the stress relaxation post-ramp.

In Ref.~\citeS{ISI:000315141600016} we showed that a state of initially
homogeneous shear will be unstable towards starting to form bands
after the ramp ends if the stress as a function of accumulated strain
just {\em before} the ramp ended had negative slope. See the dotted
line in the left part of Fig.~\ref{fig:ramp}. In other words,
criterion (\ref{eqn:startup}) applies, interpreted in the manner just
discussed.  (Caveats about the number of dynamical variables in the
generalised constitutive model used to derive the criterion do not
apply post ramp. Other, milder assumptions are discussed in
Ref.~\citeS{ISI:000315141600016,ISI:000329357400005}.)

Numerical studies of the Rolie-poly model of polymers and wormlike
micelles are consistent with this
prediction~\citeS{ISI:000329357400005,ISI:000263389500067,ISI:000319198000008}.
For a ramp rate $\gdot$ in the range $1/\taud < \gdot < 1/\taur$, the
stress shows an overshoot during the ramp at a critical strain of $O(1)$.
Provided this strain is exceeded, instability to banding will ensue
post-ramp. See the lower curve in Fig.~\ref{fig:ramp}. In contrast,
for ramp rates $\gdot > 1/\taur$ the development of chain stretch
causes linear elastic response during the ramp itself, stabilising the
system against banding immediately post-ramp. See the upper curve in
Fig.~\ref{fig:ramp}.  However, that stabilising stress then quickly
decays on a timescale of $O(\taur)$, leaving the sample in a state as if
no stretch had developed in the first place, and therefore susceptible
to banding.  (In fact, that is only true if an effect known as
`convective constraint
release'~\citeS{ISI:A1996UB61100009,ISI:000329357400004} is not too
strong.)  The Rolie-poly model thus predicts transient banding as the
sample relaxes back to equilibrium after a rapid strain ramp.  This is
consistent with early theoretical
intuition~\citeS{MarrucciGrizzuti1983a}, and with experimental
observations in polymers~\citeS{Wangetal2010a, Wangetal2009c,
  Wangetal2006a, Wangetal2007a, Fangetal2011a,
  Archeretal1995a,ISI:000289526200001,doi:10.1021/ma9004346} and
wormlike micelles~\citeS{Wangetal2008c}.

In the SGR and fluidity models of soft glasses, the stress rises
linearly during a fast strain ramp and decays relatively slowly after
it: (\ref{eqn:startup}) is not satisfied, and no banding is predicted.
(This is however still consistent with the prediction of banding in
shear startup at more modest flow rates, as discussed in
Sec.~\ref{sec:startup} above.) Indeed, this author does not know of
any experimental observations of banding after step strain in soft
glasses.

\section{Banding in perpetually time-dependent flows}
\label{sec:permanent}

%
\begin{figure*}[tbp]
\includegraphics[width=8.0cm]{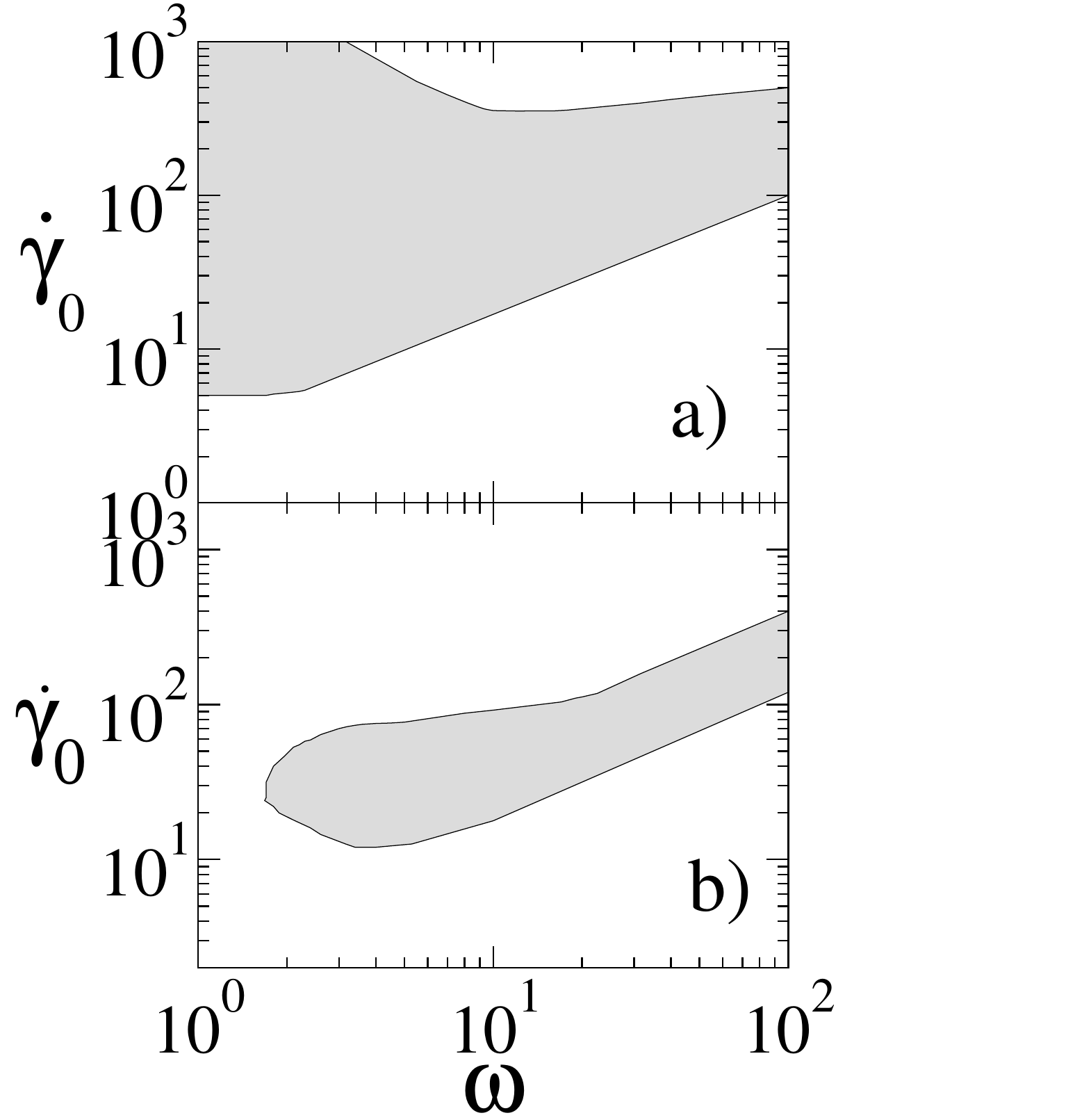}
\caption{Shear banding in LAOStrain. Shaded areas indicate the regimes
  of shear rate amplitude $\gdot_0$ and cycle frequency $\omega$ in
  which shear banding might be expected in a LAOStrain experiment with
  an imposed strain rate $\gdot(t)=\gdot_0\cos(\omega t)$, for a
  polymeric fluid with a non-monotonic underlying constitutive curve
  (panel a) and a monotonic constitutive curve (panel b). Data are
  taken from calculations in the Rolie-poly model of polymeric
  fluids~\citeS{Carter}.}
\label{fig:laos}
\end{figure*}
%

We turn now to an imposed flow that is perpetually time-dependent:
large amplitude oscillatory shear (LAOS)~\citeS{ISI:000297489400003}.
Our remarks here will be brief: a longer manuscript by this author and
coworkers has been submitted to the same issue of this
journal~\citeS{Carter}.

We focus mainly on LAOStrain with an imposed strain rate
$\gdot(t)=\gamma_0\omega\cos(\omega t) =\gdot_0\cos(\omega t)$, such
that any given experimental run is prescribed by the strain rate
amplitude $\gdot_0$ and cycle-frequency $\omega$, or equivalently the
strain amplitude $\gamma_0$ and $\omega$. (Expressed in units of the
fluid's inverse intrinsic relaxation time, $\gdot_0$ and $\omega$ are
often respectively termed the Weissenberg and Deborah number.)
Typically, after many cycles a pseudo-steady state (often called an
``alternance state'') is attained in which the fluid's response is
invariant from cycle to cycle, $t\to t+ 2n\pi/\omega$. We focus on
that regime, discarding any earlier cycles in which the response is
still settling to the flow.  Our aim is to understand in what regimes
of applied $\gdot_0$ and $\omega$ banding might arise, and to sketch
these in the plane of $\gdot_0,\omega$, noting that any coordinate
pair in this plane refers to a single LAOS experiment at the given
amplitude and frequency.  To do so, it is helpful to consider first
the dynamics of an underlying base state flow that is (artificially)
assumed to remain homogeneous.

In a LAOS experiment performed at a low frequency $\omega\to 0$, the
fluid will slowly explore its underlying stationary constitutive curve
as the strain rate sweeps progressively up and down (over both
positive and negative values) during a cycle.  In this way, the
so-called viscous Lissajous-Bowditch curve ({\it i.e.,} the stress
signal plotted parametrically as a function of strain rate round the
cycle) is expected to have the same form as the fluid's underlying
constitutive curve (Fig.~\ref{fig:steady}a,c).  If this is
non-monotonic, shear banding might then be expected in any
low-frequency LAOS experiment that has a strain rate amplitude
$\gdot_0=\gamma_0\omega$ sufficiently large to enter the banding
regime, according to the criterion (\ref{eqn:steady}) for banding in
steady shear.

At higher frequencies we might instead expect a LAOStrain experiment
to (loosely) correspond to a repeating sequence of forward and
backward shear startup runs. In any LAOS experiment of sufficiently
large strain amplitude $\gamma_0$, the elastic Lissajous-Bowditch
curve of stress plotted parametrically as a function of strain might
then be expected to show overshoots reminiscent of those in the stress
startup curve associated with a single startup run.  These overshoots
might further be expected to trigger banding in each half of the
cycle, according to a criterion resembling (\ref{eqn:startup}) for
banding in startup.  This should hold whether or not the stationary
constitutive curve that determines the fluid's response to a steady
flow (or a low-frequency LAOS run) is non-monotonic or monotonic.

This intuition was confirmed numerically in the Rolie-poly model of
polymers and wormlike micelles in
Ref.~\citeS{ISI:000263389500067,Carter}. The region of the
$\gdot_0,\omega$ plane in which shear banding was found is sketched in
Fig.~\ref{fig:laos}a) for a fluid with a non-monotonic constitutive
curve.  This shows banding at low frequency $\omega\to 0$, consistent
with the fact that such a fluid also supports banding in steady shear
flow. It also shows banding at high frequencies, reminiscent of
banding in a fast shear startup, in each half cycle of the elastic
Lissajous-Bowditch curve. For a fluid with a monotonic constitutive
curve (Fig.~\ref{fig:laos}b) the regime of low frequency banding is
absent, as expected.  Importantly, however, the regime of high
frequency banding remains. In both Figs.~\ref{fig:laos}a and b this
regime of banding eventually closes off at very high $\omega$ once the
solvent stress swamps the polymer contribution. (Data not shown.)

A thorough study of the effects of model parameter values was
conducted in Ref.~\citeS{Carter}. As expected, a stronger tendency to
banding, over larger regions of the $\gdot_0,\omega$ plane, was found
for decreasing values of solvent viscosity $\eta$, decreasing levels
of convective constraint release, and increasing entanglement number.
Indeed, moving these parameter values too far in the opposite
direction can eliminate banding. This is to be expected: a large
Newtonian viscosity swamps nonlinear viscoelasticity, for example.

While the details of Figs.~\ref{fig:laos}a,b) are likely to be model
dependent, these findings could have wider significance in suggesting
that banding might arise quite generically in flows with a strong
enough sustained time-dependence, even in fluids that do not support
bands in steady flow.

Depending on the degree of shear banding that arises, the
Lissajous-Bowditch curves of the banded flow state can differ quite
significantly from those of the base state calculated within the
assumption of a homogeneous flow. Indeed, in some cases shear bands
can persist around the entire cycle. This should lend caution to
attempts to develop rheological fingerprints within theoretical
calculations that assume the flow to remain homogeneous.  Theoretical
studies that do account for banding in LAOStrain can be found in
Refs.
~\citeS{ISI:000283686800004,ISI:000257623000062,ISI:000263389500067,Carter}.

As explored further in Ref.~\citeS{Carter}, shear banding is also
predicted to arise in polymers subject to LAOStress.

Our discussion of LAOS has so far concerned ergodic fluids such as
polymers and wormlike micelles, with finite stress relaxation
timescales. Work by Rangarajan Radhakrishnan with this author
concerning LAOStrain in the SGR model of soft glasses is also
currently under review.  As noted above, this model has a yield stress
and a constitutive curve that rises monotonically beyond it,
precluding true steady state banding.  In view of this, and the
preceding discussion, we might likewise expect the SGR model to
respond homogeneously to an imposed LAOStrain experiment at low
frequency $\omega\to 0$.

However the SGR model also displays rheological aging: in the absence
of flow its stress relaxation timescales increase as a function of the
sample age. An applied flow can then halt aging and restore an
effective sample age set by the inverse flow rate. As a result, the
response of the SGR model to a low frequency LAOStrain comprises a
complicated sequence of processes in which it alternately ages into a
solid-like state during the low shear phase of the cycle, then yields
via a stress overshoot and associated banding in the high shear phase.
In retrospect this is not surprising: an aging material has no
characteristic relaxation timescale against which to compare the
frequency $\omega$ of the applied flow.  In view of this, and more
broadly, shear banding may prove an integral feature of the response
of soft glassy materials to imposed flows of arbitrarily slow time
variation, even in the absence of true zero-frequency banding.

Experimentally, shear banding has been observed during LAOS in polymer
solutions~\citeS{tapadiaravin06,Wangetal2010a}, dense
colloids~\citeS{ISI:000242219900019}, carbon black
gels~\citeS{ISI:000342206200016,Gibaud15}, foams~\citeS{Rouyer}, non
Brownian PMMA suspensions~\citeS{ISI:000288162500028} and also in
wormlike micellar surfactants that are known to shear band in steady
state~\citeS{ISI:000303482200002,ISI:000301673400005,Wagner}.

\section{Conclusions}

In this \precis we have summarised criteria for shear banding in
steady and time-dependent flows of complex fluids and soft solids.
These criteria were derived analytically within a constitutive model
written in a highly generalised (though still differential) form, and
are supported by experimental observations, particle based
simulations, linear stability analysis and numerical solutions of
several widely used constitutive models in the exemplary contexts of
polymeric fluids (polymers and wormlike micelles) and soft glassy
materials (dense emulsions, dense colloids, microgels, {\it etc.}).
While the evidence supporting the picture presented here is therefore
quite convincing, we nonetheless now consider any caveats and
uncertainties that remain.

Our most important caveat concerns the generality of the stress overshoot
criterion for banding in shear startup.  Although indeed derived in a
constitutive model written in a generalised form, to make progress
analytically this allowed for only two dynamical viscoelastic
variables. As things stand, the status of the criterion more generally
is not completely clear.  For example, it appears not to apply in a
straightforward way in the phenomenological Giesekus model of
polymers. It does, though, convincingly hold in the Rolie-poly model
of polymers, and in the fluidity and SGR models of soft glasses. Its
verification in the SGR model seems an important result in this
context, because that model's constitutive equation is of integral
form, and so effectively has infinitely many dynamical variables.
Nonetheless, future work would be welcome to try to generalise the
criterion further, and to delineate more fully its regimes of
applicability.

The criteria put forward for the other time-dependent protocols (step
stress and during the stress relaxation following a fast strain ramp)
are not subject to any limitations concerning the number of dynamical
variables. Milder assumptions made in their derivations are discussed
in the original papers~\citeS{ISI:000315141600016,ISI:000329357400005}.

While the criteria presented predict the onset of an instability to
the formation of bands, that instability must obviously be strong
enough and persist for long enough in any time-dependent protocol to
ensure that observable bands arise before the homogeneous base state
regains stability. (Put more technically: instability is
characterised in the calculation by a positive eigenvalue, which must
remain positive and of large enough amplitude for long enough to
ensure noticeable banding~\citeS{ISI:000293534200007}.) Clearly, for the
example of startup, stronger stress overshoots are more likely to give
observable banding.  Weaker ones instead give transient instability
and enhanced spatial fluctuations, but without leading to
macroscopically observable bands (consistent with the absence of
banding altogether in the regime of slow startup flows where
overshoots are absent).

Our calculations to date have assumed the inertialess limit of
creeping flow. In this limit, the eigenvector governing the onset of
shear rate heterogeneity $\delta\gdot$ has the form $\delta \gdot =
-\delta \sigma_{v}/\eta$ where $\sigma_{v}$ is the shear component of
the viscoelastic stress, and $\eta$ is the viscosity of the
background Newtonian solvent, and/or any viscoelastic modes fast
enough not to be ascribed their own dynamical evolution.  For most
materials this background viscosity is very small, predicting a strong
degree of heterogeneity in the flow field, $\delta \gdot$, compared to
that in the viscoelastic stress, $\delta \sigma_{v}$. This predicts
potentially rather violent banding that may, ultimately, be tempered
by inertia.  While order of magnitude estimates suggest this should
not be an important effect, concrete calculations are in progress to
check this in more detail.

In polymeric fluids, numerical studies have so far mainly focused on
the Rolie-poly model~\citeS{likhtmangraham03}. This is
microscopically sophisticated enough to incorporate the dynamical
processes of reptation, chain stretch, and convective constraint
release, while also being simple enough to allow numerical progress.
However it contains only a single reptation mode and a single stretch
relaxation mode. Work is in progress to check the effects of multiple
relaxation modes, in unbreakable polymers, on the effects discussed.
In wormlike micelles (which are sometimes called ``living polymers''
due to their reversible breakage and recombination dynamics) a single
mode description should already capture most of the physics (because
breaking narrows the relaxation spectrum). Indeed, it would be
interesting to perform a comprehensive study of time-dependent flows
in wormlike micelles over the full phase diagram of concentration,
including regimes of both non-monotonic and monotonic underlying
constitutive curves.

Finally, although the effects of flow-concentration coupling are well
understood in situations of steady state
banding~\citeS{ISI:000185756400080,ISI:000183381200013,ISI:000184319400008},
their role in the time-dependent phenomena discussed above remains to
be clarified.

The author hopes that this \precis will provide a helpful guide to
situations in which shear banding might be expected in complex fluids
and soft solids subject to steady and time-dependent flows. Future
work by other authors would be welcome to verify the criteria
suggested here, to generalise them further, and/or to delineate any
regimes in which they might break down. This seems particularly
important for the case of shear startup, where it has been more
difficult than in other protocols to obtain a universal criterion free
of some caveats.

\section{Postlude: criteria for necking in extensional filament stretching}

%
%
\begin{figure}[tbp]
\includegraphics[width=7.0cm]{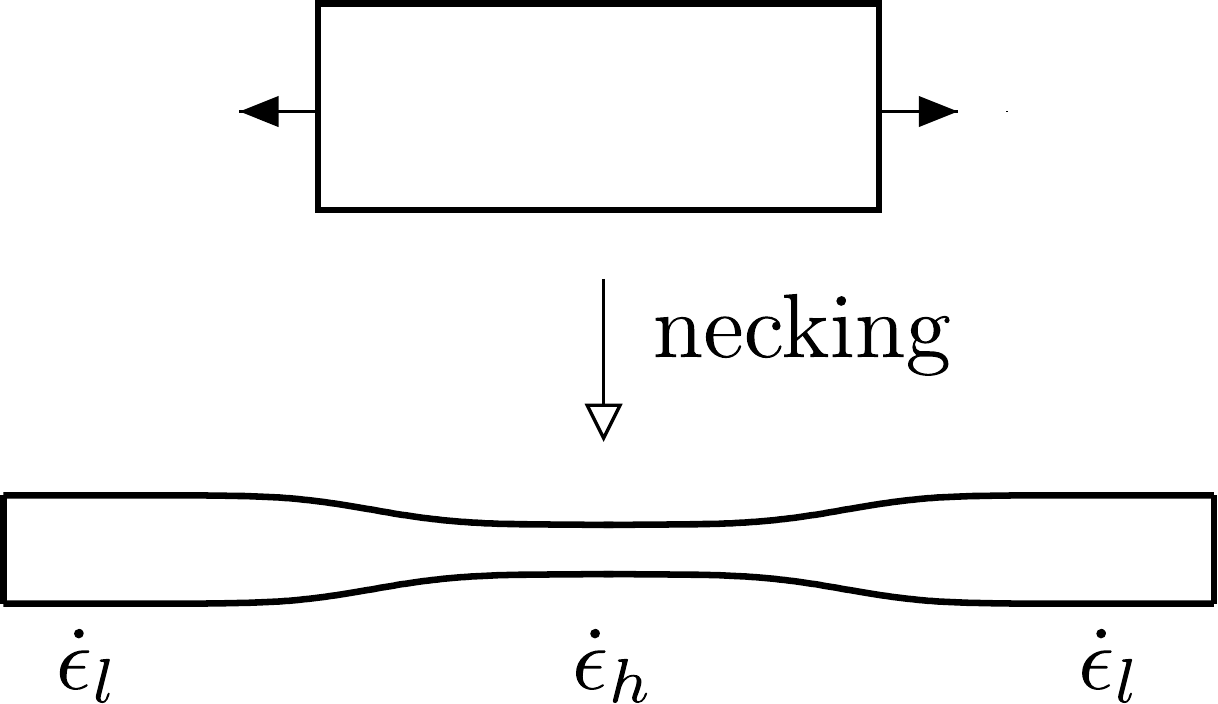}
\caption{Sketch of necking in extensional filament stretching. A
  cylindrical filament is seen here side-on.}
\label{fig:neck}
\end{figure}
%

This section summarises work by David Hoyle with this author currently
under review at the Journal of Rheology, in manuscripts ``Criteria for
extensional necking in complex fluids and soft solids: imposed Hencky
strain rate protocols'' and ``Criteria for extensional necking in
complex fluids and soft solids: imposed tensile stress and force
protocols''.

Having focused on shear flow so far, we now conclude with a brief
postlude concerning extensional flows.  In particular we consider the
phenomenon of necking in a cylindrical filament (or planar sheet)
subject to stretching, as sketched in Fig.~\ref{fig:neck}.  Here a
state of initially homogeneous flow, in which the filament is
extending and thinning in a uniform way along its length, gives way to
a heterogeneous state with a higher extension rate and more pronounced
thinning in some part of the sample. A comparison between the sketches
in Figs.~\ref{fig:steady} and~\ref{fig:neck} suggests an analogy
between shear banding and extensional necking: the relevant
deformation field (shear rate in Fig.~\ref{fig:steady} and extension
rate in Fig.~\ref{fig:neck}) becomes heterogeneous in both cases.

With that analogy in mind, in
Refs.
~\citeS{PhysRevLett.107.258301,ISI:000352990500012}, together with the
manuscripts with Hoyle under review, we developed criteria for the
onset of necking, separately for the flow protocols of step tensile
stress, step tensile force, and startup of constant Hencky strain
rate. As before, these were derived by studying the linearised
dynamics of small heterogeneous perturbations, which are the
precursors of a neck, about a state of initially homogeneous
extensional flow. Also as before, they are expressed in terms of
characteristic signatures in the shapes of the relevant underlying
rheological response function for the given protocol.  We now briefly
summarise them, referring the reader to Refs.
~\citeS{PhysRevLett.107.258301,ISI:000352990500012}, and the
manuscripts with Hoyle under review,
for a more detailed discussion, and for comprehensive citation of the
motivating literature, which is beyond the scope of this article.

Throughout we consider a highly viscoelastic filament in which bulk
stresses dominate surface tension. We also neglect flow-concentration
coupling. Also throughout, the symbol $\sigmae$ denotes the true (and
in general time-evolving) tensile stress (the time-evolving tensile
force divided by the time-evolving cross sectional area of the
filament).  It does not denote the so-called engineering tensile
stress, which is simply the tensile force divided by the (constant)
initial cross sectional area of the filament as measured at the start
of the run.

\begin{figure}[tbp]
\includegraphics[width=8.0cm]{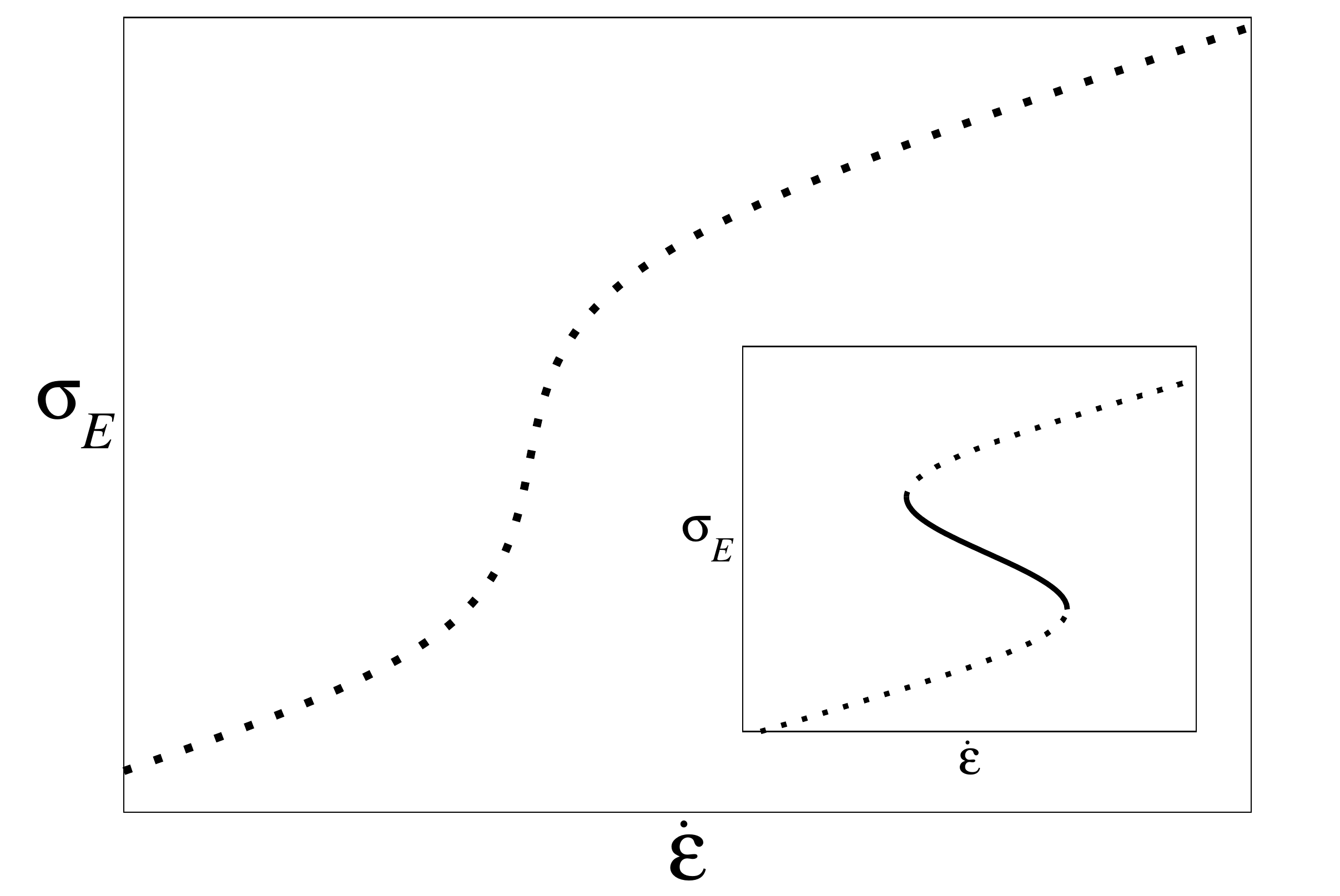}
\caption{Underlying stationary constitutive relation between tensile
  stress and Hencky strain rate, calculated within the assumption of a
  homogeneous extensional flow.  A state of initially homogeneous
  extensional flow is linearly unstable, in the regime of positive
  constitutive slope, to the formation of a neck. As described in the
  text, this result also determines the necking dynamics following the
  imposition of a tensile stress. Solid line denotes stability; dotted
  line denotes instability.}
\label{fig:extensionCC}
\end{figure}

By analogy with our discussion of shear banding above, a useful
underpinning concept is that of the stationary homogeneous
constitutive relation between the tensile stress $\sigmae$ and the
Hencky strain rate $\edot$, calculated by (artificially) assuming that
a filament can attain a state in which the stress and strain rate are
linked by this time-independent relation, with all the flow variables
remaining homogeneous along the filament. Performing a linear
stability analysis (at the level of a slender filament approximation)
for the dynamics of small heterogeneous perturbations about an
initially homogeneous and stationary state on this constitutive curve,
with the wavevector of the perturbations along the length of the
filament, then reveals instability to necking in any regime where this
constitutive curve has positive slope
\be
\frac{d\sigmae}{d\edot}>0.
\ee
This tells us that a state of initially homogeneous extensional flow,
in which the filament is drawing out and thinning in a uniform way
along its length, cannot be maintained in any regime where the
underlying extensional constitutive curve is positively sloping. See
Fig.~\ref{fig:extensionCC}.  Given that most materials indeed have a
positively sloping extensional constitutive curve, this suggests that
most materials will neck when stretched, which is indeed consistent
with experience. An interesting prediction, however, is that of
stability against necking in any regime of negative extensional
constitutive slope. See the inset of Fig.~\ref{fig:extensionCC}.  Note
the stark contrast to the corresponding result for shear banding, in
which instability is predicted for negative constitutive slope,
Eqn.~\ref{eqn:steady}.

%
%

While the calculation just discussed provides useful intuition, in
practice it is not usually possible to prepare a filament in a state
of steady uniform extensional flow on the constitutive curve because
such a flow is usually unstable, as just shown. In practice one must
compute the stability properties of a filament in which stretching was
recently commenced. We therefore now consider in turn the three common
protocols of step tensile stress, step tensile force, and startup of
constant Hencky strain rate. The results that we shall discuss were
obtained in analytical calculations performed within highly
generalised constitutive descriptions, and confirmed numerically in
several models of polymer dynamics~\citeS{Larson1988} (the Oldroyd B,
Giesekus, fene-CR, Rolie-poly~\citeS{likhtmangraham03} and
pom-pom~\citeS{McleLars98} models), and in tensorial versions of the
soft glassy rheology and fluidity models of soft
glasses~\citeS{ISI:000352990500012}.

\subsection{Step stress}

We consider first a filament subject at some time $t=0$ to the
switch-on of a constant tensile stress $\sigmae$, which is held
constant thereafter. (For times $t<0$ the sample was undeformed with
all internal stresses well relaxed.) In this case, calculations in the
polymer models listed above show that the strain rate quickly attains
its steady state value on the extensional constitutive curve before
any appreciable necking develops. The criterion
\be
\label{eqn:extStress}
\frac{d\sigmae}{d\edot}>0
\ee
for necking thereafter then applies, as sketched in
Fig.~\ref{fig:extensionCC}.

\subsection{Step force}

Consider now a filament subject at some time $t=0$ to the switch-on of
a constant tensile force $F$, which is held constant thereafter.  
In this protocol, typically, the sample attains a
state of flow on the underlying homogeneous constitutive curve, then
progressively sweeps up this curve as the stress necessarily increases
in time to maintain constant force as the cross sectional area thins.
The criterion~\ref{eqn:extStress} for the onset of necking then
applies to good approximation.

%
\begin{figure}[tbp]
\includegraphics[width=7.5cm]{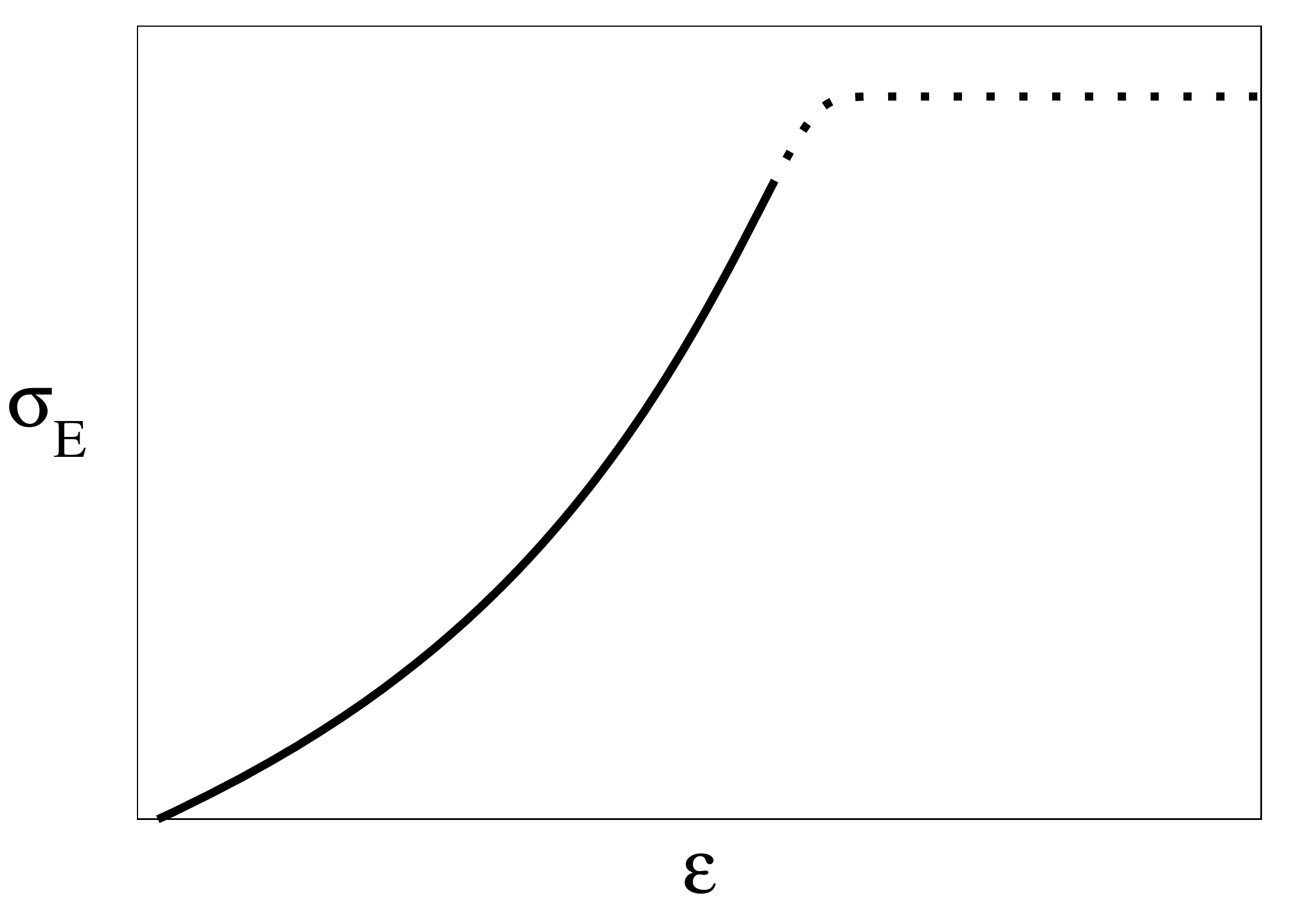}
\caption{Typical evolution of the tensile stress following the
  switch-on of a constant Hencky strain rate.  The regime of linear
  instability to necking is shown as dotted.}
\label{fig:extensionHencky}
\end{figure}
%

\subsection{Constant Hencky strain rate}

We consider finally the case of a filament that is initially at rest
and with any residual stresses well relaxed, subject at some time
$t=0$ to the switch-on of a Hencky strain rate to some value $\edot$
that is held constant thereafter.  (By this we mean that the nominal
Hencky strain rate as averaged along the filament is held constant.
Once necking arises, the true Hencky strain rate will vary along the
filament's length. As long as the filament remains uniform, however,
these nominal and true rates coincide.)  Measured in response is the
tensile stress startup signal $\sigmae(t)$ as a function of the time
$t$ (or accumulated strain $\epsilon=\edot t$) since the inception of
the flow.  In Ref.~\citeS{PhysRevLett.107.258301} we showed in the
polymer models listed above that the filament will be unstable to
necking if
\be
\frac{d^2\sigmae}{d\epsilon^2}<0,
\ee
that is, if the tensile stress shows downward curvature as a function
of the accumulated Hencky strain. See Fig.~\ref{fig:extensionHencky}.

In some models (the Rolie-poly model without chain stretch, the
pom-pom model with saturating chain stretch, and the SGR and fluidity
models) an additional mode of instability is possible, given by
\be
\label{eqn:EC}
\frac{dF_{\rm el}}{d\epsilon}<0.
\ee
(Indeed this mode can also arise, relatively rarely, under conditions
of constant imposed tensile stress or tensile force.)
This derivative needs careful
interpretation. It is calculated by evolving the full dynamics of any
model up to some strain $\epsilon$, then in the next increment of
strain over which the derivative of the tensile force $F$ is
calculated, disabling the model's relaxational dynamics and evolving
only the elastic loading terms.  As far as we are aware, this is the
closest counterpart in viscoelastic materials of the original
\considere criterion, $dF/d\epsilon<0$, for necking in
solids~\citeS{Considere}. It is important to note, however, that
(\ref{eqn:EC}) does not coincide with the original \considere
criterion, which in general fails to correctly predict the onset of
the necking instability. Indeed it is unclear whether it is possible
to access the elastic derivative of (\ref{eqn:EC}) experimentally
apart from in the limit of infinite extension rate, where relaxational
dynamics become unimportant and (\ref{eqn:EC}) simply coincides with
the original \considere criterion. (In polymer models the onset of
this mode also appears related to the presence of a very flat region
in the underlying constitutive curve at the strain rate in question,
although more work is needed to explore this suggestion fully.)
Necking in the elastic limit of viscoelastic models was also discussed
in~\cite{Hassager1998}.

It is hoped that the criteria just summarised will provide a useful
field-guide to the onset of necking instability in filament
stretching.  A fuller discussion of them can be found in
Refs.
~\citeS{ISI:000352990500012,PhysRevLett.107.258301}, together with the
manuscripts with Hoyle currently under review.

{\it Acknowledgements} The author thanks Katherine Carter, Mike Cates,
John Girkin, David Hoyle, Robyn Moorcroft, Peter Olmsted and
Rangarajan Radhakrishnan for enjoyable collaboration on these topics;
and Gareth McKinley, for suggesting the necking calculation to her.
Thanks are due to all these, and to Thibaut Divoux, for a critical
reading of the manuscript.  The author thanks Katherine Carter for
assistance in preparing Fig.~\ref{fig:laos}, and Ewan Hemingway for
assistance in preparing Fig.~\ref{fig:neck}.  Funding of the research
leading to these results was provided by the UK's EPSRC (EP/E5336X/1);
and the European Research Council under the European Union's Seventh
Framework Programme (FP/2007-2013), ERC grant agreement number 279365.

\vspace{0.5cm}

S.M.F. dedicates this manuscript to the memory of Professor Alexei
E. Likhtman.

\bibliographystyle{prsty} 

\end{document}